\documentclass[aps,twocolumn,superscriptaddress]{revtex4}

\usepackage{amsmath}
\usepackage{amssymb}
\usepackage{graphicx}
\usepackage{subfigure}
\usepackage{epstopdf}
\usepackage{amsmath}
\usepackage{bm}

\usepackage[dvipsnames]{xcolor}
\newcommand{\rev}[1]{{
		#1}}
\newcommand{\frev}[1]{{
		#1}}
\newcommand{\srev}[1]{{
		#1}}
\usepackage{ulem} % needed for \sout-command

\def\rmd{\mathrm{d}}
\def\rmi{\mathrm{i}}
\def\rme{\mathrm{e}}

\def\H{\mathcal{\widehat{H}}}

\def\D{\Delta}

\begin{document}
\bibliographystyle{apsrev}

\title{Geometry~-~dependent effects in Majorana nanowires}

\author{A.~G.~Kutlin}
\affiliation{Max Planck Institute for the Physics of Complex Systems, D-01187 Dresden, Germany}
\author{A.~S.~Mel'nikov}
\affiliation{Institute for Physics of Microstructures, Russian
Academy of Sciences, 603950 Nizhny Novgorod, GSP-105, Russia}
%\affiliation{
%Lobachevsky State University of Nizhny Novgorod, 23 Gagarina, 603950 Nizhny Novgorod, Russia}

\date{\today}

\begin{abstract}
	Starting from the Bogolubov -- de Gennes theory describing the induced p-wave superconductivity in the Majorana wire of an arbitrary shape, we predict a number of intriguing phenomena such as the geometry-dependent phase battery (or a phi-Josephson junction with the spontaneous superconducting phase difference) and generation of additional quasiparticle modes at the Fermi level with the spatial position tuned by the external magnetic field direction. This tuning can be used to extend the capabilities of the braiding protocols in Majorana networks.
\end{abstract}

\maketitle

Recent advances in the technology of the semiconducting nanowires and various hybrid superconducting structures on their basis has opened new horizons both for the study of fundamental problems of superconductivity with nontrivial topology  and the device engineering for quantum computing. All possible applications of these systems essentially exploit the topological transition in the quasiparticle spectrum and the resulting emergence of the so-called Majorana quasiparticle states with the energies inside the superconducting gap. By now, the observation of the Majorana states is reported in several systems such as the semiconducting nanowires with strong spin-orbit interaction and superconducting covering \cite{LutchynOr,OregOr,Chang,Higginbotham,Krogstrup,Albrecht}, 
\frev{Yu-Shiba-Rusinov (YSR) states in chains of magnetic adatoms \cite{ad},} 
and several others. Clearly, the application of these Majorana systems to quantum computing, quantum information, and quantum memory will demand the construction of the networks of rather complex configurations \cite{nets} which, in turn, raises a natural question about the importance of the geometry -- dependent effects (such as wire bending, turns, connections or loop formation) in the underlying physics of the induced superconductivity.
\frev{The consequences of the nanowire bending should be of particular importance for the systems with combined effects of spin-orbit coupling and magnetic field, as it has been discussed for carbon nanotubes in \cite{marcus10}}.

\srev{Before focusing on the geometry -- dependent effects specific for topologically nontrivial systems it is natural and useful to discuss briefly some basic facts related to the influence of geometry on the superconducting states with trivial topology.}
For standard s-wave superconducting wires, tapes, stripes, etc., the geometry effects have been studied for many decades in the context of different applications (see, e.g., \cite{clem11} and references therein).
\frev{From these studies we know that the bending of the superconducting wire or strip strongly affects the distribution of the superflow resulting in the strengthening of the supercurrent density near the sharp turns or corners; this strengthening originates from the hydrodynamic-type continuity equation and boundary conditions.}
Being, of course, very important for the optimization of superconducting device operation, the physics of this current redistribution is far from any issues related to the microscopic mechanism of the Cooper pair formation and their internal structure. The geometry effects can become particularly important in the case of mesoscopic superconductors of the size of several coherence lengths where they are known to be responsible for the formation of exotic vortex configurations (see \cite{mel} and references therein).

\srev{In view of strong anisotropy of superconducting correlations in the topologically nontrivial systems it is important to mention here that the geometry -- dependent effects are known to become much more interesting when one considers the anisotropic superconducting pairing, i.e., in d- or p-wave superconductors with the gap nodes at the Fermi surface.} 
Indeed, the change in the geometry of the sample boundary in this case may affect the scattering of the quasiparticle trajectories by coupling the directions of electron momenta with different signs of the gap. Clearly, in such a way the modification of the boundary geometry affects the subgap quasiparticle states bound to the surface as well as the gap itself even without any applied transport current (see, e.g., \cite{dahm}). The bending of a superconducting wire can even generate a nontrivial flux pattern \cite{sigrist14,sigrist18} or spatial pattern of the spin triplet correlations \cite{ying}.

In the systems with induced superconducting order such as Majorana nanocircuits one has to separate the influence of the geometry on the superconducting order parameter in the topologically trivial and nontrivial phases. Indeed, in the former case the geometry effects should have either standard hydrodynamic nature or result from the inverse proximity phenomenon modified in the presence of the strong Zeeman (or exchange) and spin-orbit effects. This modification has been recently shown to result in the generation of spontaneous currents and described successfully within the generalized Ginzburg -- Landau (GL) theory \cite{buzdin1,balatsky,mironov17,buzdin2} written for the s-wave order parameter. On the other hand, the induced superconducting correlations in the topologically nontrivial phase are known \cite{LutchynOr,OregOr} to be triggered to the effective triplet p-wave state. We show that this topological transition and the emergent superconducting order are strongly affected by the system geometry. 
\srev{Thus, the main goal of our work is to study the geometry -- dependent effects specific for the topologically nontrivial superconducting states and illustrate these geometric effects by the particular examples of experimentally measurable characteristics of Majorana nanowires.}

\begin{figure}[!t]
	\center{\includegraphics[width=0.45\textwidth]
		{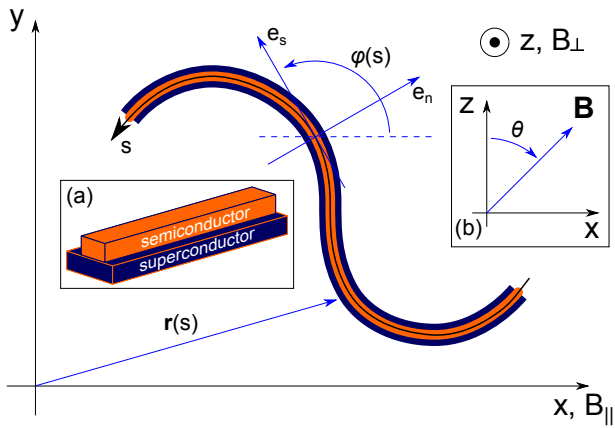}}
	\caption{
		\label{fig:hs} Illustration of the coordinate system used throughout the work.
		Inserts: a) heterostructure configuration; b) magnetic field direction
}
\end{figure}

In order to elucidate our main results we, first, consider a general form of the p-wave gap
$p_{\gamma} = p_x + \rmi \gamma p_y
$ confirmed further by microscopical calculations.
Here $\gamma$ is a model dependent constant governing chiral
$p_\pm =p_x \pm \rmi p_y
$ ($\gamma=\pm 1$) or nodal
$p_0 = p_x
$ ($\gamma=0$) structure of the gap.
Let's consider a planar curved wire situated in the $XY$-plane (see Fig.~\ref{fig:hs}).
The order parameter
$p_{\gamma} = p_s (\cos\varphi(s) \pm \rmi \gamma \sin\varphi(s))
$
written in terms of the momentum component $p_s$ along the wire and the local angle $\varphi(s)$ of the wire is strongly affected by the sample geometry both in chiral and nodal cases. Here the coordinate $s$ parameterizes the wire location
$\bm{r}=(x(s),y(s))
$, Fig.~\ref{fig:hs}.
Indeed, in the chiral case
$\gamma=\pm 1$, $p_{\gamma}
$ reveal itself as
the fully-gapped order parameter
$p_{\pm}=p_s \rme^{ \pm \rmi \varphi(s)}
$  with
the imbedded spatially inhomogeneous phase structure $\varphi(s)$.
%and the nonzero internal orbital momentum.
Moreover, the finite curvature of the circuit in this case induces a nonzero geometry dependent gain of superconducting phase along the Majorana wire. The gradient of this geometrical phase $\varphi(s)$ should excite the spontaneous current in the wire which in turn causes the screening current in the superconductor in the opposite direction~\footnote{It is interesting to note that this mechanism of generation of the spontaneous currents has an analogy in the GL-type consideration \cite{balatsky} describing the inverse proximity effect at the surface of the superconductor in contact with the finite size ferromagnet. Choosing the magnetic moment perpendicular to the surface and introducing the spin-orbit coupling into this problem one can get the surface GL term proportional to the scalar product of the superconducting and magnetization currents. Clearly, a ferromagnetic strip placed on the S surface and forming, e.g., a ring should in this case generate a spontaneous circulating supercurrent because of the different lengths of the inner and outer ring radii.}. In the nodal case, $\gamma=0$, the gap
$p_{\gamma} = p_s \cos\varphi(s)
$ is modulated along the wire and, thus, may host zero energy modes depending on the magnetic field direction. Microscopically in the realistic nanocircuits the value of the $\gamma$ parameter and, thus, the structure of the superconducting phase is expected to be controlled by the orientation of the external magnetic field or the exchange field in the magnetic adatoms. In particular, the field perpendicular to the plane $XY$ should generate the chiral superconducting gap
$\propto \rme^{ \pm \rmi \varphi(s)}
$ with the nonzero internal orbital momentum. Tilting the magnetic field with respect to the plane $XY$ leads to the induced order parameter with a modulated amplitude, which can cause strong changes in the quasiparticle spectra including the subgap quasiparticle states. The rest part of the paper aims to confirm the above qualitative arguments by more solid microscopic considerations based on the Bogoliubov -- de Gennes (BdG) theory.

Further we discuss the basic equations describing a model heterostructure consisting of the s-wave superconductor and a planar curved semiconducting wire with a strong SOC (see Fig.\ref{fig:hs}). Assuming the external magnetic field $\bf{B}$ to be homogenous, we generalize standard BdG equations describing the quasiparticle wavefunctions in a semiconducting wire with the induced superconductivity, strong Zeeman field and spin-orbit coupling \cite{LutchynOr,OregOr} for the case of a curved wire.
\srev{Indeed, neglecting the wire's width compared to the local radius of curvature, we can project the full Rashba Hamiltonian to the lowest transverse mode
$R_0(r_\perp)$ looking for the solution in the form $\Psi_{0n}(r_\perp,s)=R_0(r_\perp)\Psi_n(s)$, as it was done in Ref.~\cite{meijer02} for a semiconducting ring. This approximation implies that the confining energy scale of the transverse modes well exceeds all the other energy scales in the problem. Hence, the resulting BdG Hamiltonian for $\Psi_n(s)$ takes the form}
\begin{eqnarray}
\H_{smc} =
    \left[ \begin{array}{*{2}{c}}
		\H_e & \D_{ind} \\
		\D_{ind}^* & -\sigma_y \H_e^*\sigma_y
	\end{array} \right],
\label{eq:genProxHam1}
\\
\H_e = \frac{\hat{P}^2}{2m} -\mu - \frac{u}{2} \{ \hat{\sigma}_n(s), \hat{P} \} - g\mu_B\bf{B}\cdot\hat{\sigma},
\label{eq:genProxHam}
\end{eqnarray}
where $\hat{P}=\hat p - e A_{s}/c$ is a gauge-invariant kinematic momentum along the curve, $\hat{p} =- \rmi\hbar\partial_s$ is a canonical momentum, $A_{s}$ is a component of the vector potential $\bm{A}$ along the wire, \rev{$\mu$ is a chemical potential,} $u$ is a Rashba spin-orbit coupling constant, $\mu_B$ is the Bohr magneton, $g$ is the Lande factor which is known to be large for the systems under consideration,
$\hat\sigma =(\sigma_x,\sigma_y,\sigma_z)
$ is a vector of Pauli matrices, and $\hat{\sigma}_n(s) = \sigma_x\sin\varphi(s)- \sigma_y \cos\varphi(s)$ is its component in the sample $XY$-plane perpendicular to the wire. The curly brackets denote the anti-commutator of two operators. For the particular case of the ring the effective 1D Hamiltonian (\ref{eq:genProxHam}) can be reduced to the one obtained in Ref.~\cite{meijer02} assuming the function $\varphi(s)$ to be linear.

Note that, to some extent, the above model can be also applied to the analysis of a one-dimensional set of coupled YSR states induced in the superconductor in the presence of the  chain of magnetic adatoms. Indeed, in a limit of a dense chain with the intercite spacing $a$ much less than the coherence length $\xi$ of the host superconductor, the adatom chain term $a J_0 \sum_j (\bm{S}_j \cdot \hat{\sigma}) \delta(\bm{r-R}_j)$ turns into the inhomogeneous Zeeman term. The latter problem can be reduced \cite{unitar,ad} to the one with a homogeneous magnetic field of strength $J_0/2$ and a corresponding SOC term with $\hat{\sigma}_n(s)$ determined not by a geometry of a wire but by a texture of the spin $\bm{S}_j$.

Most of the previous theoretical works in the field have been focused on the topologically nontrivial phenomena described by the Hamiltonian (\ref{eq:genProxHam}) in the case of a straight wire. These phenomena are known to reveal themselves in the limit of large Zeeman energy $g\mu_B B > \sqrt{|\D_{ind}|^2 +\mu^2}$ when the wire hosts Majorana states at its ends \cite{oppen14}. In this regime the spin-dependent terms in the above Hamiltonian cause the generation of the effective p - wave superconducting correlations. The latter fact can be easily illustrated in the so-called Kitaev limit, i.e., in the limit of small SOC energy compared to Zeeman field. This condition allows to treat the spin-orbit coupling perturbatively and project the $4 \times 4$ Bogoliubov-de-Gennes Hamiltonian (\ref{eq:genProxHam})  to the lower Zeeman band (for detailed derivation see Appendix \ref{app:kitaev}).
As a result, we obtain the following effective low-energy Bogoliubov-de-Gennes Hamiltonian:
\begin{eqnarray}
\H_{eff} =
	\left[ \begin{array}{*{2}{c}}
		\frac{1}{2m}\left(\hat{p} - \frac{e}{c}\tilde{A}_s \right)^2 - \tilde\mu & \frac{1}{2p_F}\{\D,\hat{p}\} \\
		\frac{1}{2p_F}\{\D^*,\hat{p}\} & -\frac{1}{2m}\left(\hat{p} + \frac{e}{c}\tilde{A}_s \right)^2 + \tilde\mu
	\end{array} \right],
\label{eq:eff_ham}	
\\
\D = \rmi\frac{u p_F \D_{ind}(s)}{g \mu_B B}
\Big(
		\cos\varphi(s) -\rmi \cos \theta \sin\varphi(s)
\Big),
\label{eq:gapoperator}
\end{eqnarray}
where $p_F$ is the Fermi momentum, 
\srev{$\tilde{A}_s$ and $\tilde\mu$ are the vector and chemical potentials renormalized in accordance with the Appendix \ref{app:kitaev},} 
and the magnetic field is chosen as $\bm{B}=B [\sin\theta,0,\cos\theta]$, see Fig.\ref{fig:hs}. Note that the form of the off-diagonal gap operator in Eq.~(\ref{eq:gapoperator}) agrees well with the qualitative considerations in the introduction, with $\gamma = -\cos\theta$.

\srev{
The phase $\varphi_{ind}$ of the induced gap $\D_{ind}$ is imposed by the phase $\varphi_{sc}$ of the superconductor's s-wave order parameter $\D_{sc}$ which in turn is influenced by the coupling to the semiconducting wire through the inverse proximity effect. Following the works \cite{mironov17, balatsky, buzdin2}, one can estimate the consequences of the latter effect using the GL-like model with the surface terms $\alpha_{soc} [ \bm{e}_z, \bm{B}] \cdot (\D_{sc}^* \hat{\bm{D}} \D_{sc} + h.c.)$ and $\alpha_\varkappa [ \bm{\varkappa}, \bm{B}] \cdot (\D_{sc}^* \hat{\bm{D}} \D_{sc} + h.c.)$, where $\hat{\bm{D}} = -\rmi \hbar \nabla + 2 e \bm{A}/c$, $\bm{e}_z = [ \bm{e}_n, \bm{e}_s]$, $\bm{\varkappa} = \bm{e}_n \partial_s \varphi(s)$, and the parameters $\alpha_{soc}$ and $\alpha_\varkappa$ are the phenomenological parameters proportional to the small probability of electron tunnelling between the wire and the superconductor. The surface terms generate in the s-wave superconductor spontaneous inhomogeneous phase gains $\varphi_{soc}$ and $\varphi_\varkappa$. The phase gradient along the wire $\partial_s \varphi_{soc} \propto \alpha_{soc} [ \bm{e}_z, \bm{B}] \cdot \bm{e}_s$ appears to be proportional to the additional term $m u c/e * \sin\theta\sin\varphi(s)$ in the renormalized vector potential $\tilde{A}_s$ obtained in the Kitaev limit in the Appendix \ref{app:kitaev}. In its turn, the phase $\varphi_\varkappa$ is proportional to the geometrical phase in the p-wave gap $\D$ given by the Eq.~\ref{eq:gapoperator}. Thus, due to the aforementioned relations between $\varphi_{ind}$, $\varphi_{sc}$, $\varphi_{soc}$ and $\varphi_\varkappa$, the inverse proximity effect can be responsible for the partial compensation of the phase gradient and the vector potential in the semiconducting wire. This partial compensation occurs in the surface  layer of the thickness smaller than the superconducting coherence length and finally for large GL length scales gives us the spontaneous phases determined by the small parameters $\alpha_{soc}$ and $\alpha_\varkappa$. Thus, inside the wire the total superfluid velocity is also proportional to these parameters. Hereafter, we will omit the small Doppler shift of the quasiparticle energy determined by these effects.}

The resulting chirality of the induced order parameter should reveal itself, e.g., in a wire forming a closed loop. Indeed, the wire curvature causes the appearance of an additional gain of the phase of the gap function along this loop. Taking the magnetic field to be perpendicular to the plane of the wire, one can easily find that the additional phase gain in the semiconductor with respect to the superconductor is equal to $\pm 2\pi$ for $\theta=\pi$ and $\theta=0$, correspondingly. As a result, the winding number of the induced p-wave order parameter differs from the one in the primary superconducting loop by $\pm 1$. In other words, keeping in mind the obvious fact that the magnetic field introduces a certain number $N$ of vortices inside the primary superconducting ring one can see that the vorticity number for the semiconducting ring with induced superconductivity is $N\pm 1$.

\begin{figure}[!htb]
	\center{\includegraphics[width=0.4\textwidth]
		{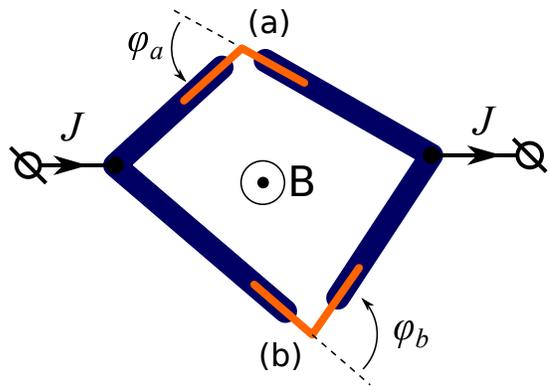}}
	\caption{
		\label{fig:squid} Possible experimental setup for detection of the spontaneous superconducting phase induced by the wire curvature.
	}
\end{figure}

\srev{The above consideration allows us to suggest an interesting experiment aimed at the detection of the geometry controlled superconducting phase in topologically non-trivial systems and based on the measurement of a nonzero spontaneous phase difference in the ground state of a Josephson junction.}
Note that the Josephson devices with a nonzero phase offset  (known also as phase batteries) are in the focus of current research interest \cite{buzdin2,exp,nazarov,bergeret,houzet} due to their perspective importance for the field of quantum computing. Experimentally, the characteristics of the geometry-controlled phase battery can be probed using various standard SQUID setups, e.g., one can consider a dc SQUID with two one-channel Josephson junctions shown in Fig.~\ref{fig:squid}. Indeed, a total flux through the SQUID is expressed in terms of the s-wave phase differences at the junctions, whereas the current through each junction must be calculated using an actual phase difference of the induced gap. As a result, for a total current through the SQUID we obtain
\begin{eqnarray}
	J = J_a(\theta_a + \varphi_a) + J_b(\theta_b - \varphi_b),
\end{eqnarray}
where $\varphi_{a,b}$ are the geometrical phase gains at the junctions $(a)$ and $(b)$, and the s-wave phase differences $\theta_{a,b}$ are related as
\begin{eqnarray}
	\theta_a-\theta_b = \frac{2\pi \Phi}{\Phi_0},
\end{eqnarray}
with $\Phi$ and $\Phi_0 = \pi \hbar c/e$ staying for the total flux through the SQUID and the flux quantum, correspondingly. Note that we do not assume the phase batteries $(a)$ and $(b)$ to have the same current-phase relations and use different functions $J_a(\phi)$ and $J_b(\phi)$ instead. Irrespective to the particular form of these functions the resulting shift of the critical current dependence vs. magnetic flux is described by the expression
\begin{eqnarray}
	J_c=J_c \left( \Phi + \frac{\varphi_a + \varphi_b}{2\pi} \Phi_0 \right) \ .
	\label{eq:squid}
\end{eqnarray}
A particular form of the Josephson current-phase relation can be obtained following the line of derivation in \cite{beenakker91}. For example, assuming the junctions to be equally fabricated ($J_a(\phi) \equiv J_b(\phi) \equiv J_0(\phi)$), in the zero temperature limit for a transparent single channel junction ($J_0(\phi) \propto \sin(\phi/2) \mathrm{sign}( \cos (\phi/2) )$) we find
\begin{eqnarray}
J_c \propto 1 + \Bigg| \cos\left( \frac{\pi\Phi}{\Phi_0} + \frac{\varphi_a + \varphi_b}{2} \right) \Bigg|,
\end{eqnarray}
in agreement with the previous considerations. The general result given by the Eq.~\ref{eq:squid} also holds for a SQUID with an arbitrary number of different junctions; in such cases $\varphi_a + \varphi_b$ must be replaced by a sum of all the corresponding geometrical phase differences.

In another generic limit of the in-plane magnetic field, $\theta=\pi/2$, the nodal gap $\D=\D_0\cos \varphi(s)$ turns to zero when $\varphi = \pm\pi/2$; such points separate the regions with different signs of the gap function and can be viewed as Andreev domain walls which are known to host the zero energy quasiparticle bound states.

To get the explicit solution in the quasi-classical approximation we
\srev{look for the eigenfunctions of the Hamiltonian (\ref{eq:eff_ham}) in the form $\psi^\alpha(s) \rme^{ \alpha \rmi k_F s}$ with $\psi^\alpha(s) = \left[u^\alpha(s), v^\alpha(s)\right]$ being a two-component wave function envelope, $\alpha=\pm$ corresponding to two opposite momenta of the quasi-classical modes and $k_F$ standing for the Fermi wave vector. Assuming $k_F^{-1}$ to be negligible compared to the radius $R$ of the wire's local curvature we obtain}
the resulting quasiclassical BdG equations in the form
\begin{eqnarray}
- \rmi \hbar v_F \frac{\rmd}{\rmd s} \psi^\alpha(s)=( \alpha E^\alpha \tau_z - \rmi \D(s) \tau_x) \psi^\alpha(s),
\label{eq:in-plane}
\end{eqnarray}
where $v_F$ is the Fermi velocity, $\tau_{x,z}$ are Pauli matrices acting in the electron-hole space, and $\psi^\alpha(s) = \left[u^\alpha(s), v^\alpha(s)\right]$ is the two component quasiclassical wavefunction. Hereafter we choose the gauge with a positive gap amplitude
\begin{eqnarray}
\D_0 = \frac{u p_F |\D_{ind}|}{g \mu_B B}.
\end{eqnarray}

\begin{figure}[!htb]
	\center{\includegraphics[width=0.4\textwidth]
		{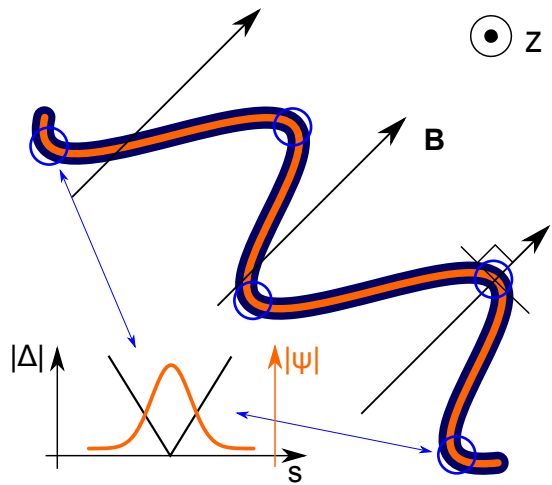}}
	\caption{
		\label{fig:bound_states} Domain walls and bound states in the nodal case.
	}
\end{figure}

Let the one of the aforementioned Andreev domain walls to be located at $s=0$; then, in a vicinity of the domain wall, where $\D(s) \propto s$, the problem is described by the equations similar to the standard harmonic oscillator equations, and we get the spectrum in the form (cf. \cite{osc}):
\begin{eqnarray}
(E_n^\alpha)^2 = 2 \D_0^2 \frac{\xi_0}{R} n,
\end{eqnarray}
where
%$R$ is a radius of the local curvature and
$\xi_0~=~\hbar v_F/\D_0$ is the coherence length in the semiconductor. This spectrum clearly allows the zero-energy modes localized at the domain walls (see Fig.~\ref{fig:bound_states}) with the wavefunction amplitudes being proportional to $\exp(- s^2/2 \xi_0 R)$ shown in the inset.
\frev{It is worth mentioning that, in the quasiclassical limit, there are always two noninteracting bound states at each domain wall; these states differ by the direction of the running phase in the exponents $\exp(\pm \rmi k_F s)$.}
Thus, in the system with $N$ domain walls we have $2N$ (or, in a case of the wire with open ends, $2N+2$) localized quasiparticle states close to the Fermi level. The corresponding degeneracy of these levels is removed by 
\srev{non-quasi-classical corrections, scattering on sharp inhomogeneities, Doppler shifts due to incomplete phase compensation, and by} 
the exponentially small overlapping of the quasi-classical wave functions.

To find the splitting of these modes 
\srev{due to the overlapping} 
we consider a general solution of Eq.~(\ref{eq:in-plane})
\begin{eqnarray}
\psi^\alpha(s)= 	\exp\left( \tau_x K(s) \right) U^\alpha(s,0)	\psi^\alpha(0),
\label{eq:gen_sol}
\end{eqnarray}
where
\begin{eqnarray}
K(s) =\int_{0}^{s} \frac{ \D(s')}{\D_0} \frac{\rmd s'}{\xi_0},
\end{eqnarray}
and $U^\alpha(s,0)$ is a series in the powers of energy $E^\alpha$:
\begin{eqnarray}
\begin{split}
U^\alpha(s,0) = 1 + \int_{0}^{s}\rmd s' \frac{\rmi \alpha E^\alpha}{\hbar v_F}\tau_z\rme^{2 \tau_x K(s')} \Big( 1 +
\\
+\int_{0}^{s'}\rmd s'' \frac{\rmi \alpha E^\alpha}{\hbar v_F} \tau_z\rme^{2 \tau_x K(s'')} \big( 1+...\big)\Big).
\end{split}
\label{eq:U}
\end{eqnarray}
Using these expressions and taking, for example, periodic boundary conditions, one can find the quasiclassical spectrum from the condition
\begin{eqnarray}
\det\left[\rme^{ \tau_x K(L)} U^\alpha(L,0) - \rme^{-\rmi \alpha k_F L} \right]=0,
\label{eq:loop}
\end{eqnarray}
where $L$ is the length of the loop and $k_F$ is the Fermi wave vector. Restricting ourselves by the second order terms in the series (\ref{eq:U}) we obtain:
\begin{eqnarray}
(E_{gs}^\alpha)^2 = 2 \D_0^2 A \left( \cosh \left( K(L) \right)  - \cos(k_F L)\right), \label{eq:loop_gse}
\\
A = \left(  c_+ \rme^{ K(L) } + c_- \rme^{ - K(L) } \right)^{-1},
\\
c_\pm = \xi_0^{-2} \int_0^{L} \rmd s' \int_0^{s'} \rmd s'' \rme^{\pm 2 \left(K(s'')-K(s')\right)}.
\end{eqnarray}

As it can be seen from Eq.~(\ref{eq:gen_sol}), the corresponding wave functions look like two sets of quasiparticles localized either at the odd or even domain walls. Indeed, in the zeroth order in tunnelling between the domain walls, the wave functions in the vicinity of odd and even domain walls appear to be orthogonal. Taking $\D(s) \propto \cos(\varphi(s))$ one obtains $K(L) = 0$
and, thus,  Eq.~(\ref{eq:loop_gse}) can be simplified even further and written as
\begin{eqnarray}
(E_{gs}^\alpha)^2 = 4 \D_0^2  B \sin^2(k_F L/2) \ ,
\\
B = \left( \xi_0^{-2} \int_0^L \int_0^L \rme^{2 \left(K(s'') - K(s') \right) } \rmd s' \rmd s'' \right)^{-1} \ .
\end{eqnarray}
In the case of a ring with a small curvature ($ R \gg \xi_0$) the expression takes the form
\begin{eqnarray}
(E_{gs}^\alpha)^2 =  \frac{4 \D_0^2 \xi_0}{\pi R} \sin^2 (\pi R k_F ) \rme^{- 4 R / \xi_0 },
\end{eqnarray}
which looks quite similar to the spectrum of coupled Majorana modes \cite{majorana-pair}. Note, finally, that changing the orientation of the magnetic field in the plane $XY$ we can change the position of the states bound to the Andreev domain walls tuning, thus, the spectrum of the low energy quasiparticle modes; the situation is somewhat similar to the one in the two-dimensional p-wave superconducting disks\cite{guinea17}.

The presence of the additional zero-energy modes localized at the aforementioned domain walls can strongly affect the coupling between the end Majorana states of the open wire. The corresponding corrections should be important for the structure of the ground-state wave function which can, thus, be tuned by the magnetic field rotation. As a result, we get an alternative way of manipulating this ground-state wave function by adiabatic moving of the domain walls controlled by the magnetic field rotation. 
\srev{The corresponding unitary transformation of the wave function can extend the capabilities of the existing Coulomb-mediated braiding protocols in Majorana networks (see e.g. \cite{karzig17} and references therein)}.

To summarize, we have shown that changing the geometry of the Majorana 1D systems (such as nanowires or chains of magnetic adatoms) one can effectively tune the quasiparticle spectrum inducing the spontaneous order parameter phase texture or the low energy quasiparticle modes. Thus, our consideration provides a new recipe to manipulate the ground-state wave function and the characteristics of the Josephson links with an arbitrary ground state phase difference in the Majorana networks.

The authors would like to thank A.~I.~Buzdin, S.~V.~Mironov, and I.~M.~Khaymovich for the fruitful discussions. This work was supported, in part, by the Russian Foundation for Basic Research under Grants Nos. 17-52-12044 and 19-31-51019, German Research Foundation (DFG) Grant No. KH 425/1-1, by the Foundation for the Advancement to Theoretical Physics and Mathematics BASIS Grant No. 17-11-109 and by the German-Russian Interdisciplinary Science Center Grant No. F-2019b-9\_d. In the part concerning the analysis of the phase batteries, the work was supported by the Russian Science Foundation (Grant No. 17-12-01383). A.S.M.  appreciates warm hospitality of the Max-Planck Institute for the Physics of Complex Systems, Dresden, Germany, extended to him during the visits when this work was done.

\appendix

\section{The Kitaev limit \label{app:kitaev}}
To get the Hamiltonian (\ref{eq:eff_ham}) from Eqs.(\ref{eq:genProxHam1})-(\ref{eq:genProxHam}), let's write all four BdG equations explicitly 
\begin{eqnarray}
\begin{cases}
\left( \hat{\xi} - Z_z  - E \right) u_\uparrow =
\left(\hat{S} + Z_x \right) u_\downarrow - \Delta_{ind} v_\downarrow
\\
-\left( \hat{S}^\dagger + Z_x \right) u_\uparrow + \Delta_{ind} v_\uparrow =
\left( E - \hat{\xi} - Z_z  \right) u_\downarrow
\\
\Delta_{ind}^* u_\uparrow -\left( \hat{S}^{*\dagger} + Z_x \right) v_\uparrow =
\left( E + \hat{\xi}^* + Z_z \right) v_\downarrow
\\
\left( -\hat{\xi}^* + Z_z  - E \right) v_\uparrow =
-\Delta_{ind}^* u_\downarrow +\left( \hat{S}^* + Z_x \right) v_\downarrow
\end{cases},
\end{eqnarray}
where $Z_{x,z}(\theta)=g \mu_B B_{x,z}(\theta)$ are the components of the Zeeman term, $\hat{S}=\rmi u\{\rme^{-\rmi \varphi(s)},\hat{P}\}/2$ originates from the SOC term, $\hat{\xi} = \hat{P}^2/2m - \mu_0$ is a kinetic term, $u_{\uparrow\downarrow}(s)$ and $v_{\uparrow\downarrow}(s)$ are the spin-up and spin-down electron and hole components of the eigenfunctions, and $E$ is the state's energy. 
As a next step, we need to eliminate the off-diagonal contributions coming from the Zeeman interaction term by the following rotation of variables:
\begin{eqnarray}
\begin{cases}
u_- = \cos\frac{\theta}{2} u_\uparrow + \sin\frac{\theta}{2} u_\downarrow \ ,
\\
u_+ = -\sin\frac{\theta}{2} u_\uparrow + \cos\frac{\theta}{2} u_\downarrow \ ,
\\
v_+ = \cos\frac{\theta}{2} v_\downarrow + \sin\frac{\theta}{2} v_\uparrow \ ,
\\
v_- = -\sin\frac{\theta}{2} v_\downarrow + \cos\frac{\theta}{2} v_\uparrow \ .
\end{cases}
\end{eqnarray}
In the new variables $u_\pm$ and $v_\pm$ we obtain:
\begin{eqnarray}
\begin{cases}
\left( \hat{\xi} - Z - \hat{S}_\parallel  - E \right) u_- =
\hat{S}_\perp u_+ - \Delta_{ind} v_+
\\
-\hat{S}_\perp^\dagger u_- + \Delta_{ind} v_- =
\left( E - \hat{\xi} - Z -  \hat{S}_\parallel \right) u_+
\\
\Delta_{ind}^* u_- - \hat{S}_\perp^{*\dagger} v_- =
\left( E + \hat{\xi}^* + Z + \hat{S}_\parallel^* \right) v_+
\\
\left( -\hat{\xi}^* + Z  - E + \hat{S}_\parallel^* \right) v_- =
-\Delta_{ind}^* u_+ + \hat{S}_\perp^* v_+
\end{cases},
\end{eqnarray}
where
\begin{eqnarray}
\hat{S}_\parallel = \frac{u \sin\theta }{2} \Big\{\sin\varphi(s),\hat{P}\Big\},
\\
\hat{S}_\perp = \rmi \frac{u}{2} \Big\{ \cos\varphi(s) - \rmi \cos\theta\sin\varphi(s),\hat{P} \Big\}
\end{eqnarray}
are the SOC components in the new basis and $Z=g\mu_B B$ is the Zeeman energy. Excluding now the high-energy components $u_+(s)$ and $v_+(s)$, we can write the equations describing the behavior of the low-energy components as follows:
\begin{eqnarray}
\begin{cases}
\left( \hat{\xi} - Z - \hat{S}_\parallel  - E \right) u_- = \hat{T}_{ee}u_- + \hat{T}_{eh} v_-
\\
\left( -\hat{\xi}^* + Z  - E + \hat{S}_\parallel^* \right) v_- = \hat{T}_{he} u_- + \hat{T}_{hh} v_-
\label{eq:A6}
\end{cases}.
\end{eqnarray}
Here
\begin{eqnarray}
\begin{split}
\hat{T}_{ee} = - \hat{S}_\perp &\left( E - \hat{\xi} - Z -  \hat{S}_\parallel \right)^{-1} \hat{S}_\perp^\dagger
\\
& -\Delta_{ind} \left( E + \hat{\xi}^* + Z + \hat{S}_\parallel^* \right)^{-1} \Delta_{ind}^*
\end{split},
\\
\begin{split}
\hat{T}_{hh} = - \Delta_{ind}^* &\left( E - \hat{\xi} - Z -  \hat{S}_\parallel \right)^{-1} \Delta_{ind}
\\
& -\hat{S}_\perp^* \left( E + \hat{\xi}^* + Z + \hat{S}_\parallel^* \right)^{-1} \hat{S}_\perp^{*\dagger}
\end{split},
\\
\begin{split}
\hat{T}_{eh} = \hat{T}_{he}^\dagger = &\hat{S}_\perp \left( E - \hat{\xi} - Z -  \hat{S}_\parallel \right)^{-1} \Delta_{ind}
\\
&+ \Delta_{ind} \left( E + \hat{\xi}^* + Z + \hat{S}_\parallel^* \right)^{-1} \hat{S}_\perp^{*\dagger}
\end{split}
\end{eqnarray}
are the operators describing the scattering of the low-energy modes at the high-energy ones. 
Generalizing the perturbative procedure described in \cite{oppen14} for the case of the inhomogeneous angle $\varphi (s)$
we omit the second order corrections in the operators $\hat{S}$ and $\Delta_{ind}$ and obtain the off-diagonal terms of the Hamiltonian (\ref{eq:eff_ham}).
Indeed, the diagonal terms $\hat{T}_{ee}$ and $\hat{T}_{hh}$ are proportional to the squares of the operators $\hat{S}$ and $\Delta_{ind}$ and, hence, can be neglected, while $\hat{T}_{eh}$ can be rewritten in the form
\begin{eqnarray}
\hat{T}_{eh} \simeq \frac{1}{2Z} \left( \Delta_{ind} \hat{S}_\perp^{*\dagger} - \hat{S}_\perp\Delta_{ind} \right).
\end{eqnarray}
After some algebra, we obtain
\begin{eqnarray}
\hat{T}_{eh} \simeq
-\left\{ \rmi\frac{ u \Delta_{ind}}{2 g\mu_B B}
\Big( \cos\varphi(s) - \rmi \cos\theta\sin\varphi(s) \Big), \hat{p} \right\},
\end{eqnarray}
i.e. precisely the upper off-diagonal term of the Hamiltonian~(\ref{eq:eff_ham}). Finally, redefining the vector and chemical potentials in the diagonal entries of the system (\ref{eq:A6}) to get rid of the terms $\hat S_\parallel$ and $Z$ correspondingly as
\begin{eqnarray}
	\tilde{A}_s = A_s + \frac{m u c \sin\theta}{e} \sin\varphi(s),
	\\
	\tilde\mu = \mu + Z + \frac{m u^2}{2} \sin^2\theta \sin^2\varphi(s),
\end{eqnarray}
we arrive at the effective Hamiltonian in the form~(\ref{eq:eff_ham}).

\end{document}